\documentclass{article}

\usepackage[utf8]{inputenc} %
\usepackage[T1]{fontenc}    %
\usepackage[colorlinks=false,linkcolor=black,anchorcolor=black,citecolor=blue,filecolor=black,menucolor=black,runcolor=black,urlcolor=black]{hyperref}       %
\usepackage{url}            %
\usepackage{booktabs}       %
\usepackage{amsfonts}       %
\usepackage{nicefrac}       %
\usepackage{microtype}     %
\usepackage{calc} %
\usepackage{xcolor}         %
\usepackage{bm}         %
\usepackage{graphicx}  		%
\usepackage{amsmath} %
\usepackage{cite} %
\usepackage{diagbox}
\usepackage{siunitx}
\usepackage{multirow}
\usepackage{tabularx}
\usepackage{hhline}
\usepackage{arydshln}		%
\usepackage{mathtools} %
\usepackage{amsthm}			%
\usepackage{amssymb}			%
\usepackage{algorithm}
\usepackage{pifont}
\usepackage{threeparttable}
\usepackage{fullpage}
\usepackage[letterpaper,
            left=1.67cm,
            right=1.67cm,
            top=1.8cm,
            bottom=1.69in]{geometry}
\usepackage{outlines}

\usepackage{algpseudocode}	
\usepackage{arydshln}
\usepackage[export]{adjustbox}

\usepackage{mathtools} 
\usepackage{microtype} 
\usepackage{multirow}
\usepackage{nicefrac}       
\usepackage{pifont}   

\usepackage{url}
\usepackage{siunitx}
\usepackage{tabularx}
\usepackage{threeparttable} 
\usepackage{caption}
\usepackage{tabularray}
\UseTblrLibrary{booktabs}
\usepackage{wrapfig}
\usepackage{lipsum}
\usepackage{setspace}
\usepackage{tikz}

\usepackage{graphicx}
\usepackage{soul}          %

\definecolor{lightgreen}{rgb}{.85,1,.85}
\definecolor{lightred}{rgb}{1,.85,.85}
\definecolor{lightblue}{rgb}{.85,.85,1}
\definecolor{pink}{HTML}{EB346F}
\hypersetup{linkcolor=black, citecolor=black, filecolor=black, urlcolor=pink}
\newcommand{\hlgreen}[1]{{\sethlcolor{lightgreen}\hl{#1}}}

\newcommand{\hlblue}[1]{{\sethlcolor{lightblue}\hl{#1}}}
\newcommand{\hlbest}[1]{%
	{\colorbox{lightgreen}{\makebox(22,6){\textbf{#1}}}}
}
\newcommand{\hlsecondbest}[1]{%
	{\colorbox{lightblue}{\makebox(22,6){#1}}}
}

\renewcommand{\vec}[1]{\bm{#1}}

\def\x{\vec{x}}  %
\def\y{\vec{y}}  %
\def\z{\vec{z}}  %
\def\w{\vec{w}}  %

\def\s{\vec{s}}  %

\newcommand{\normpdf}{{\mathcal{N}}}

\theoremstyle{plain} %

\def\argmin{\mathop{\mathsf{arg\,min}}} %

\title{Plug-and-Play Priors as a Score-Based Method}
\date{}

\author{
Chicago~Y.~Park$^\dagger$,
Yuyang~Hu$^\dagger$,
Michael~T.~McCann$^\ddagger$, \\
Cristina~Garcia-Cardona$^\ddagger$, Brendt~Wohlberg$^\ddagger$, and Ulugbek~S.~Kamilov$^\dagger$\\
\small $^\dagger$Washington University in St. Louis, $^\ddagger$Los Alamos National Laboratory\\
\small \texttt{\{chicago, h.yuyang, kamilov\}@wustl.edu},\\
\small \texttt{\{mccann,  cgarciac, brendt\}@lanl.gov}
}

\begin{document}

\maketitle

\vspace{-2em} %
\begin{abstract}
\medskip\noindent
Plug-and-play (PnP) methods are extensively used for solving imaging inverse problems by integrating physical measurement models with pre-trained deep denoisers as priors. Score-based diffusion models (SBMs) have recently emerged as a powerful framework for image generation by training deep denoisers to represent the score of the image prior. While both PnP and SBMs use deep denoisers, the score-based nature of PnP is unexplored in the literature due to its distinct origins rooted in proximal optimization. This letter introduces a novel view of PnP as a score-based method, a perspective that enables the re-use of powerful SBMs within classical PnP algorithms without retraining. We present a set of mathematical relationships for adapting popular SBMs as priors within PnP. We show that this approach enables a direct comparison between PnP and SBM-based reconstruction methods using the same neural network as the prior. Code is available at \href{https://github.com/wustl-cig/score_pnp}{\textcolor{magenta}{\text{https://github.com/wustl-cig/score\_pnp}}}.

\end{abstract}

\section{Introduction}
\label{sec:Introduction}
The recovery of an unknown image $\x \in \mathbb{R}^n$ from noisy measurements $\y \in \mathbb{R}^m$ is a fundamental problem in computational imaging.
This task is commonly formulated as an optimization problem
\begin{equation} \label{eq:begin_inv_problem}
    \widehat{\x} \in \argmin_{\x \in \mathbb{R}^n} \; g(\x) + h(\x) \;,
\end{equation}
where $g(\x)$ is a data-fidelity term quantifying consistency with the observed measurements $\y$ to the unknown signal $\x$,
and $h(\x)$ is a regularizer imposing prior knowledge on $\x$.
The formulation in \eqref{eq:begin_inv_problem} corresponds to the maximum a posteriori probability (MAP) estimator when
\[
    g(\x) = -\log \, p(\y|\x) \text{ and } h(\x) = -\log \,p(\x) \;,
\]
where $p(\y|\x)$ denotes the likelihood that relates  $\x$ to the measurements $\y$, and $p(\x)$ represents the prior distribution.
For linear inverse problems of form \( \y = \vec{A}\x + \vec{e} \), where \( \vec{A} \in \mathbb{R}^{m \times n} \) represents the measurement operator and \( \vec{e} \in \mathbb{R}^n \) is the additive white Gaussian noise (AWGN), the data-fidelity term becomes \( g(\x) = \frac{1}{2} \|\y - \vec{A}\x\|^2_2 \).

Proximal algorithms~\cite{Parikh.Boyd2014} are commonly used to solve the optimization problems of form \eqref{eq:begin_inv_problem} when the functions $g$ or $h$ are nonsmooth by leveraging a mathematical concept known as the \textit{proximal operator}
\begin{equation} \label{eq:prox}
\mathrm{prox}_{\gamma h}(\z) := \argmin_{\x \in \mathbb{R}^n} \left\{\frac{1}{2}\|\x - \z\|_2^2 + \gamma h(\x) \right\} \;,
\end{equation}
where $\gamma \geq 0$ is an adjustable penalty parameter. Plug-and-play (PnP) priors is based on the observation that the proximal operator in \eqref{eq:prox} can be interpreted as a MAP denoiser for AWGN. For example, the popular PnP-ADMM~\cite{Venkatakrishnan.etal2013, Chan.etal2016} algorithm is obtained by replacing $\mathrm{prox}_{\gamma h}$ in ADMM with a more general image denoiser, $\mathsf{D}_{\sigma}$, resulting in the iterates
\begin{subequations} \label{eq:admm}
\begin{align}
    \label{subeq:admm1}
    \x_k &\leftarrow \mathrm{prox}_{\gamma g} \left( \z_{k-1} - \s_{k-1} \right)\\
    \label{subeq:admm2}
    \z_k &\leftarrow \mathsf{D}_\sigma \left( \x_k + \s_{k-1} \right)\\
    \label{subeq:admm3}
    \s_k &\leftarrow \s_{k-1} + \x_k - \z_k \;,
\end{align}
\end{subequations}
where $\sigma \geq 0$ controls the denoising strength.

Score-based diffusion models (SBMs) are powerful methods for generating samples from complex high-dimensional distributions~\cite{song2019generative, ho_NEURIPS2020_ddpm, song2021sde}. SBMs rely on the concept of a \emph{score}---%
the gradient of the log-probability of a noisy version of the desired distribution, $\nabla \log p_{\sigma}(\x)$---%
to iteratively refine samples by reducing noise over successive steps. There has also been a growing interest in using SBMs to solve inverse problems by integrating the data-fidelity term into the diffusion process~\cite{graikos2022diff_as_pnp,chung2023dps, liu2023dolce}. 

While denoising is central to both PnP and SBMs, the score-based nature of PnP is often overlooked in the literature. We argue that by interpreting PnP as a score-based method, state-of-the-art neural networks pre-trained as SBMs can serve as priors within traditional PnP methods, such as PnP-ADMM. This approach offers several advantages: (a) improved PnP performance through the use of more powerful, open-source priors, (b) flexibility to use PnP without the need for inverse stochastic differential equation solvers of SBMs, and (c) the ability to make direct, fair comparisons between PnP and score-based posterior sampling methods by using the same neural network as the prior. To this end, we present a set of mathematical relationships that enable a simple adaptation of the neural networks approximating the score in SBMs as priors within PnP. Our numerical results compare the performance of several classical PnP methods, such as DPIR, PnP-ADMM, and RED, using the same SBM neural network as a prior.

\begin{figure*}
\begin{minipage}[t]{0.495\textwidth} %
\setstretch{1.0} %
\begin{algorithm}[H] %
\caption{PnP-ADMM with score-based denoising}
\begin{algorithmic}[1]
\Require $\z_0, \s_0 = \vec{0}, \gamma > 0, \{\sigma_k\}^{K}_{k=1}$
\For{$k = 1 \text{ to } K$}
    \State \(\x_k \leftarrow \operatorname{prox}_{\gamma g}(\z_{k-1}-\s_{k-1})\)
    \begin{tikzpicture}[remember picture,overlay]
        \node[xshift=3.9cm,yshift=-0.585cm] at (0,0){%
    \includegraphics[width=2\textwidth]{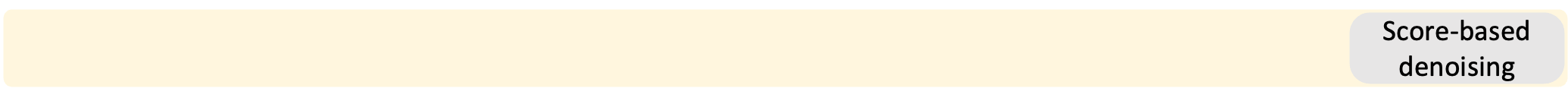}};
    \end{tikzpicture}
    \State \((c, t) \leftarrow \mathsf{param\_matching}(\sigma_k)\)
    \State \(\z_k \leftarrow (\x_k + \s_{k-1}) + c\sigma_{k}^2 \s_{\theta}(c (\x_k + \s_{k-1}),t)\)
    \State \(\s_k \leftarrow \s_{k-1} + \x_k - \z_k\)
\EndFor
\State \textbf{return} $\s_{K}$
\end{algorithmic}
\label{alg:score_based_pnpadmm}
\end{algorithm}
\end{minipage}
\hfill %
\begin{minipage}[t]{0.495\textwidth} %
\setstretch{1.0} %
\begin{algorithm}[H] %
\caption{RED with score-based denoising}
\begin{algorithmic}[1]
\Require $\s_0, \gamma > 0, \{\sigma_k\}^{K}_{k=1}$
\For{$k = 1 \text{ to } K$}
    \State \(\x_k \leftarrow \s_{k-1} - \gamma\nabla g(\s_{k-1})\)
    \State \((c, t) \leftarrow \mathsf{param\_matching}(\sigma_k)\)
    \State \(\z_{k} \leftarrow \x_{k-1} + c\sigma_{k}^2 \s_{\theta}(c \x_{k-1},t)\)
    \State \(\s_k \leftarrow \x_k - \gamma\tau (\s_{k-1}-\z_{k})\)
\EndFor
\State \textbf{return} $\s_{K}$
\end{algorithmic}
\label{alg:score_based_red}
\end{algorithm}
\end{minipage}
\end{figure*}

\section{Background}

\medskip\noindent
\textbf{PnP reconstruction.} 
There is a vast literature on different variants of PnP (see \cite{Kamilov.etal2023} for a review), including regularization by denoising (RED)\cite{romano2017RED}, deep plug-and-play image restoration (DPIR)\cite{zhang2021dpir}, and numerous others~\cite{Kamilov.etal2017, sun2019onlinePnP, ahmad2020pnpmri, sun2021scalablePNP}. RED~\cite{romano2017RED} is a popular variant of PnP that uses the denoiser residual to approximate the gradient of the implicit PnP regularizer. The steepest descent variant of RED can be expressed as
\begin{equation} \label{eq:red}
    \begin{aligned}
    \x_k &\leftarrow \s_{k-1}-\gamma\nabla g(\s_{k-1})\\
    \z_k &\leftarrow \mathsf{D}_\sigma \left( \x_k \right)\\
    \s_k &\leftarrow \x_{k} - \gamma\tau(\s_{k-1}-\z_{k}) \;,
    \end{aligned}
\end{equation} 
where $\sigma > 0$ controls the denoising strength and $\tau > 0$ is the regularization parameter. DPIR~\cite{zhang2021dpir} incorporates the DRUNet denoiser as a prior within the half-quadratic splitting (HQS) iterations using decreasing noise levels
\begin{subequations} \label{eq:hqs}
\begin{align}
    \label{subeq:hqs1}
    \x_k &\leftarrow \mathrm{prox}_{\gamma_{k}g} \left( \z_{k-1}\right)\\
    \label{subeq:hqs2}
    \z_k &\leftarrow \mathsf{D}_{\sigma_k}(\x_{k}) \;,
\end{align}
\end{subequations}
where $\sigma_k$ denotes the noise level at the $k$-th iteration and the penalty parameter $\gamma_k$ is defined as $\gamma_k = \sigma_k^2/\lambda$, with $\lambda$ being a tunable hyperparameter. 

\medskip\noindent
\textbf{PnP diffusion models.}
Given the success of SBMs in generating high-quality images and the flexibility of PnP methods to incorporate data-fidelity terms, prior works have explored PnP sampling from the posterior distribution $p(\x | \y)$ to solve inverse problems~\cite{laumont2022pnpula, bouman2023generativePnP, zhu2023DiffPIR, sun2024provablePMC, wu2024principledPnP, xu2024provably_dps_pnp}.
It is worth emphasizing that these approaches are distinct from classical PnP algorithms that seek to minimize the objective in \eqref{eq:begin_inv_problem}.
For example, denoising diffusion models for plug-and-play image restoration (DiffPIR)~\cite{zhu2023DiffPIR} extends DPIR~\cite{zhang2021dpir} by incorporating the data-consistency step directly into the sampling process as follows
\begin{align*}
    \widehat{\x}_0^{(t)} &\leftarrow \argmin_{\z \in \mathbb{R}^n} \left\{\frac{1}{2\sigma_{t+1}^2}\|\z - \x_{t+1}\|_2^2 + \log p(\z)\right\} \\
    \x_{t} &\leftarrow \mathrm{prox}_{\gamma_{k} g} \left( \widehat{\x}_0^{(t)}\right) \; + \sigma_{t}\vec{\epsilon} \;,
\end{align*}
where $\bm{\epsilon} \sim \normpdf(\vec{0}, \vec{I})$, the diffusion model is used in the first step, and $\sigma_t$ is the noise level at time $t$. Posterior sampling versions of the PnP and RED, leveraging expressive score-based generative priors and Langevin diffusion, have also been proposed~\cite{sun2024provablePMC}.

\section{Score Adaptation for PnP}

We propose a method for exploiting pre-trained SBM neural networks
as denoisers within classical PnP algorithms. This approach enables minimizing the objective with a score-based regularizer and data fidelity term without requiring reverse diffusion iterations. Additionally, the resulting PnP algorithms inherit all the theoretical convergence guarantees developed for the PnP framework.

\medskip\noindent
\textbf{Relating Score to Denoising.} Tweedie's formula~\cite{efron2011tweedie} provides a direct relationship between the score function and the MMSE denoiser for a given noise distribution.
We present a general template that enables to re-characterize any pre-trained score function as a PnP denoiser using Tweedie's formula, regardless of the noise perturbation scheme used for the training the score.
To this end, we define the following general noise perturbation scheme, encompassing all possible score training configurations
\begin{equation}\label{eq:general_noise_addition}
    \x_{c\sigma} = c(\x + \w) 
    \quad
    \w \sim \normpdf(\vec{0}, \sigma^2\vec{I}) \;,
\end{equation}
where $c$ is the scaling factor, and $\sigma$ is noise level.
The MMSE denoiser $\mathsf{D}_\sigma$ for the denoising problem~\eqref{eq:general_noise_addition}, can be related to the score function of $\x_{c\sigma}$---%
the scaled and noisy observation---%
through the Tweedie's formula. This relationship allows us to derive a general score-based denoising template
\begin{equation}\label{eq:general_tweedie}
\mathsf{D}_{\sigma}(\x)= \x + c\sigma^2 \nabla \log p_{c \sigma}(c \x) \;,
\end{equation}
where $\x \in \mathbb{R}^d$, and $p_{c \sigma}(\x)$ is the probability density function of the scaled noisy observation $\x_{c\sigma}$. Note that as $c \rightarrow 1$ and $\sigma \rightarrow 0$, the distribution $p_{c \sigma}(c \x)$ approaches the distribution of noise-free images $p(\x)$.

We will next directly apply our general denoising template to two popular classes of diffusion models.

\medskip\noindent
\textbf{Variance-Exploding SBMs.}
The variance-exploding (VE) diffusion models~\cite{song2021sde} is an important class of diffusion models, trained using the noise corruption process
\begin{equation}\label{eq:ve_noise_addition}
    \x_{\sigma_t} = \x + \w 
    \quad
    \w \sim \normpdf(\vec{0}, \sigma_{t}^2\vec{I}) \;.
\end{equation}
Comparing with the general noise perturbation in \eqref{eq:general_noise_addition}, the VE noise perturbation corresponds to the case where $c=1$ and $\sigma = \sigma_t$.
We then can map the pre-trained VE diffusion model to the PnP denoiser using \eqref{eq:general_tweedie} as
\begin{equation} \label{eq:ve_mmse_score}
\begin{aligned}
    \mathsf{D}_{\sigma}(\x) &= \x + \sigma_{t}^2 \s^{\text{VE}}_{\theta}(\x, t) \;,
\end{aligned}
\end{equation}
where $\s^{\text{VE}}_{\theta}(\x, t)$ is the noise-conditional VE score network that approximates $\nabla \log p_{\sigma_t}(\x)$.
 The conditional input $t$ corresponds to the index of the noise level $\sigma_t$ in the predefined noise sequence $\{\sigma_t\}_{t=1}^{T}$ used during training, which is explicitly defined and fixed.

\medskip\noindent
\textbf{Variance-Preserving SBMs.}
The variance-preserving (VP) diffusion model~\cite{ho_NEURIPS2020_ddpm} is another important class, where the noise perturbation keeps the total variance of the image constant throughout the noise addition steps $t$ in \([1, T]\). The VP degradation process is often expressed as
\begin{equation}\label{eq:vp_noise_addition}
    \x_{\bar{\alpha}_t} = \sqrt{\bar{\alpha}_t}\left(\x + \sqrt{\frac{1-\bar{\alpha}_t}{\bar{\alpha}_t}}\bm{\epsilon}\right)
    \quad
    \bm{\epsilon} \sim \normpdf(\vec{0}, \vec{I}) \;,
\end{equation}
where $\bar{\alpha}_t = \prod_{s=1}^t \alpha_s$, and $\alpha_t$ is chosen to ensure $\x_{\bar{\alpha}_0}$ follows desired probability distribution and $\x_{\bar{\alpha}_{T}}$ follows a known distribution such as a standard Gaussian.
Comparing with the general noise perturbation in \eqref{eq:general_noise_addition}, VP noise perturbation corresponds to the case where $c=\sqrt{\bar{\alpha}_t}$ and $\sigma = \sqrt{\frac{1-\bar{\alpha}_t}{\bar{\alpha}_t}}$. We can map the VP diffusion model to PnP denoising using our template as
\begin{equation}\label{eq:vp_general_matching}
\begin{aligned}
\mathsf{D}_{\sigma}(\x) &= \x + \frac{1 - \bar{\alpha}_{t}}{\sqrt{\bar{\alpha}_{t}}} \s^{\text{VP}}_{\theta}(\sqrt{\bar{\alpha}_{t}} \, \x, t) \;,
\end{aligned}
\end{equation}
where $\s^{\text{VP}}_{\theta}(\x, t)$ is the time-conditional VP score network that approximates $\nabla \log p_{\sqrt{\bar{\alpha}_{t}}\sigma}(\x)$ and $t \in [1, T]$ parameterizes the noise level.
Unlike VE, VP implicitly defines $t$ corresponding to the noise level through the cumulative product of $\alpha_t$ rather than an explicitly fixed noise sequence.

\medskip\noindent
\textbf{Parameter Matching.} Algorithms \ref{alg:score_based_pnpadmm} and \ref{alg:score_based_red} summarize the implementations of PnP-ADMM and RED using SBM priors, respectively. The key step in both is the SBM adaptation function $\mathsf{param\_matching}(\sigma)$, which returns the constant scaling factor \( c \) and the conditional time-step \( t \) given a desired noise level \(\sigma\) for PnP algorithms. The constant $c$ is determined by comparing the noise perturbation used in score training with the general noise perturbation formulation in \eqref{eq:general_noise_addition} under the same noise level \(\sigma\). %

Since diffusion model training discretizes the inherently continuous relationship between noise levels and time steps over a predefined time range, a specific noise level $\sigma$ cannot always be exactly matched to a particular time-step 
$t$. We address this issue by interpolating the corresponding noise sequences. For example, we first linearly interpolate the noise sequence 
\(\{\sigma_t\}_{t=1}^{T}\)  over an extended range \([1, T']\) (where \(T' > T\)) to approximate continuous mapping.
Given a noise level \(\sigma\), we identify the corresponding time step \(t'\) by finding the index in the interpolated sequence where the value is closest to \(\sigma\).
Then, the conditional time input for the VP score network is computed as 
\(T(\frac{t'}{T'})\).
A related concept was introduced in \cite{park2024randomwalks} within the context of noise-level-interpolated diffusion sampling.

\medskip\noindent
\textbf{Theoretical Convergence.} Since the SBM-based denoiser in~\eqref{eq:general_noise_addition} corresponds to the MMSE denoiser, the convergence of the corresponding PnP algorithms can be directly established using the classical results from the PnP literature~\cite{Xu.etal2020, Hurault.etal2022a, shoushtari2024priormismatch}.

\begin{figure*}[t]
\begin{center}
\includegraphics[width=\textwidth]{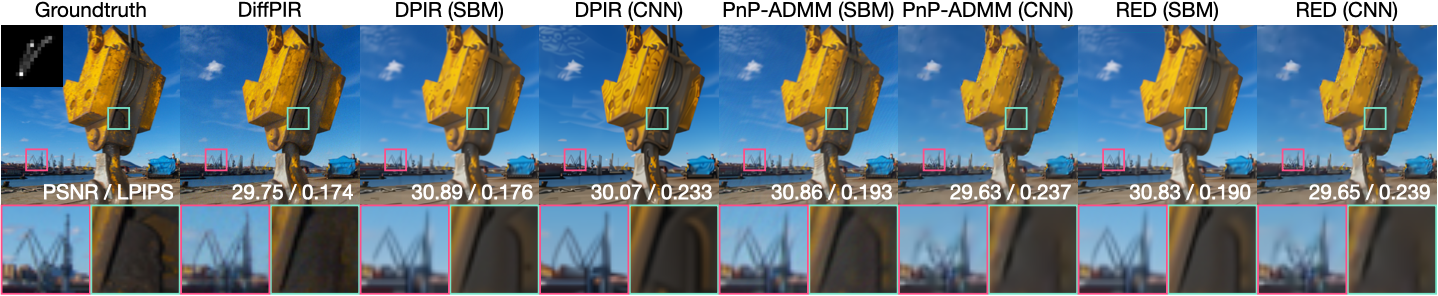}
\end{center}
\caption{Visual comparison of three classical PnP methods for motion deblurring on color images. DPIR, PnP-ADMM, and RED are compared with CNN-based and SBM-based priors~\cite{ho_NEURIPS2020_ddpm}. The figure also includes the results of DiffPIR using the same SBM as the prior. Note how our framework enables the direct comparison of DiffPIR with three classical PnP methods using exactly the same neural network as the prior.
}
\vspace{-.3cm}
\label{fig:visual_comparison}
\end{figure*}

\section{Numerical evaluation}
\label{sec:num_eval}

\medskip\noindent
\textbf{Experimental setup.}  We evaluate our algorithm on motion deblurring using 100 randomly selected images from the ImageNet test set~\cite{deng2009imagenet}, as provided in DiffPIR~\cite{zhu2023DiffPIR}.
Two motion blur kernels from \cite{Levin.etal2009} are used for the evaluations and presented in Table \ref{table:comparison_table}.

Our experiments consider three classical optimization-based PnP methods for regularized image reconstruction, including DPIR, PnP-ADMM, and RED. We additionally provide results for DiffPIR, which is the recent score-based posterior sampling method for drawing samples from the posterior distribution.

We consider three pre-trained models: DnCNN~\cite{zhang2017dncnn} for PnP-ADMM and RED, DRUNet~\cite{zhang2021dpir} for DPIR, and a pre-trained VP diffusion model~\cite{ho_NEURIPS2020_ddpm} for all PnP algorithms and DiffPIR. The DnCNN model, obtained from the DeepInv repository~\cite{tachella2023deepinverse}, was trained on ImageNet~\cite{deng2009imagenet} with a noise level of $2/255$.
We adopt the VP diffusion model from \cite{dhariwal2021beat}, trained on the ImageNet dataset.
We used pre-trained DRUNet~\cite{zhang2021dpir} that was trained across noise levels in $[0, 50/255]$ using the BSD400~\cite{martin2001CBSD68}, Waterloo Exploration~\cite{ma2016waterloodata}, DIV2K~\cite{agustsson2017DIV2K}, and Flick2K~\cite{lim2017Flick2K} datasets.

We set the number of iterations to 100 for all PnP algorithms and identified the optimal parameter choice for each using grid search. The search was conducted to optimize PSNR, SSIM, and LPIPS~\cite{zhang2018lpips} on five randomly selected ImageNet samples excluded from the test set.
The evaluation was conducted on a motion deblurring task with a measurement noise level of \(0.02\). For each PnP approach using classical CNN denoisers for regularized reconstruction, we set parameters as follows: for DPIR, \((\gamma_k = \sigma_k^2/0.27, \sigma_1 = 49 / 255, \sigma_K = 5 / 255)\), where the noise sequence \(\sigma_k\) follows a log-scale progression, starting at the maximum noise level \(\sigma_1\) and ending at the minimum noise level \(\sigma_K\); for RED, \((\gamma = 0.91, \tau = 1.1, \sigma = 2 / 255)\); and for PnP-ADMM, \((\gamma = 0.97, \sigma = 2 / 255)\).
For regularized reconstruction using SBM-based denoisers, the parameters were set as follows: for DPIR, \((\gamma_k = \sigma_k^2/0.27, \sigma_1 = 130 / 255, \sigma_K = 3 / 255)\); for RED, \((\gamma = 0.28, \tau = 3.57, \sigma = 5 / 255)\); and for PnP-ADMM, \((\gamma = 0.43 / \sigma_k^2, \sigma_1 = 120 / 255, \sigma_K = 10 / 255)\).
For the posterior sampling algorithm DiffPIR, the parameters were set as \((\lambda = 3, \zeta = 0.9)\).

Table \ref{table:comparison_table} demonstrates that substituting CNN-based priors with SBMs in DPIR, PnP-ADMM, and RED can improve standard distortion metrics (PSNR and SSIM) as well as the perception-oriented metric (LPIPS).
The comparison between the score-based posterior sampling method (DiffPIR) and other PnP methods using the same SBM prior shows that PnP methods can achieve better standard distortion metrics and comparable perception-oriented performance to DiffPIR. Figure \ref{fig:visual_comparison} illustrates that PnP algorithms with SBM denoisers improve feature recovery compared to classical approaches.

\begin{table}[t]
    \centering
    \caption{\small Quantitative evaluation of image deblurring across four setups for each PnP method. The denoiser types are indicated in parentheses: PnP methods using DnCNN and DRUNet represent classical approaches, while those using SBM denote PnP using a VP diffusion model as the denoiser.
    \hlgreen{\textbf{Best values}} and \hlblue{second-best values} are color-coded for each metric.}
    \renewcommand{\arraystretch}{0.8}
    \begin{tabular}{@{}p{3.5cm}p{0.5cm}p{0.5cm}p{0.5cm}@{}p{0.2cm}@{}p{0.5cm}p{0.5cm}p{0.5cm}@{}}
    \toprule
    & \multicolumn{3}{c}{\includegraphics[width=1.5cm]{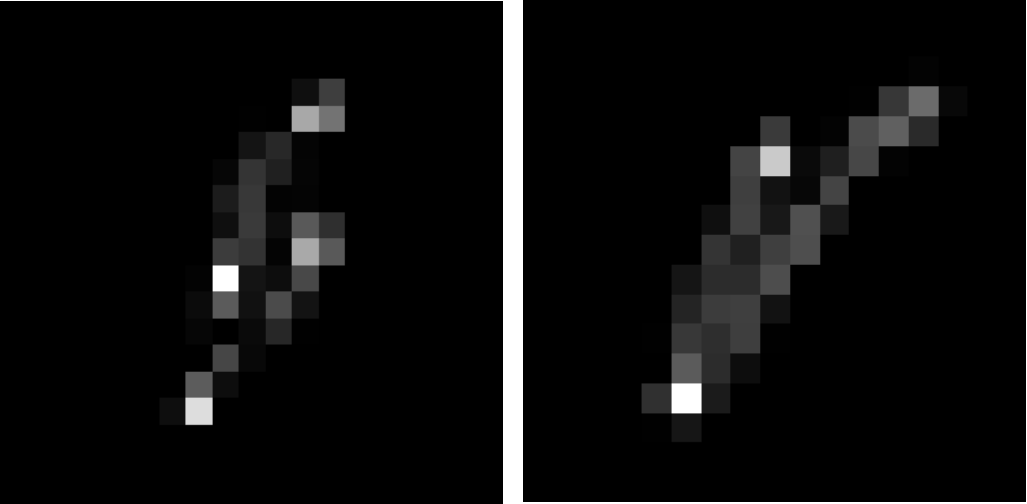}} \\[+1ex]
    & \multicolumn{1}{c}{\textbf{PSNR}$\uparrow$} & \multicolumn{1}{c}{\textbf{SSIM}$\uparrow$} & \multicolumn{1}{c}{\textbf{LPIPS}$\downarrow$} \\
    \cmidrule{1-4}\\[-1.9ex]
    Measurement  & \multicolumn{1}{c}{21.99} & \multicolumn{1}{c}{0.496} & \multicolumn{1}{c}{0.483} \\[+.5ex] \cdashline{1-4} \\[-1.2ex]
    DiffPIR (SBM) & \multicolumn{1}{c}{29.35} & \multicolumn{1}{c}{0.816} & \multicolumn{1}{c}{\hlbest{\textbf{0.190}}} \\[+.5ex] \cdashline{1-4} \\[-1.2ex]
    RED (DnCNN) & \multicolumn{1}{c}{29.21} & \multicolumn{1}{c}{0.837} & \multicolumn{1}{c}{0.279} \\[+1.0ex]
    RED (SBM) & \multicolumn{1}{c}{30.35} & \multicolumn{1}{c}{0.859} & \multicolumn{1}{c}{0.228} \\[+.5ex] \cdashline{1-4} \\[-1.2ex]
    PnP-ADMM (DnCNN) & \multicolumn{1}{c}{29.20} & \multicolumn{1}{c}{0.839} & \multicolumn{1}{c}{0.275} \\[+1.0ex]
    PnP-ADMM (SBM) & \multicolumn{1}{c}{\hlsecondbest{30.58}} & \multicolumn{1}{c}{0.857} & \multicolumn{1}{c}{\hlsecondbest{0.195}} \\[+.5ex] \cdashline{1-4} \\[-1.2ex]
    DPIR (DRUNet) & \multicolumn{1}{c}{30.52} & \multicolumn{1}{c}{\hlbest{\textbf{0.871}}} & \multicolumn{1}{c}{0.251}  \\ [+1.0ex]
    DPIR (SBM) & \multicolumn{1}{c}{\hlbest{\textbf{30.65}}} & \multicolumn{1}{c}{\hlsecondbest{0.870}} & \multicolumn{1}{c}{0.206}  \\
    \bottomrule
    \end{tabular}
    \label{table:comparison_table}
    \vspace{-0.5cm}
\end{table}

\section{Conclusion}

There has been a growing interest in score-based diffusion models for computational imaging. This letter showed that any score function used in a diffusion model can be incorporated as the prior in a classical PnP method without retraining. This perspective can lead to better PnP performance due to the use of more powerful, open-source priors available online. It additionally enables direct and fair comparisons between classical PnP methods with more recent score-based posterior sampling methods on the same task using the same neural networks as priors.

\section{Acknowledgment}
Research presented in this article was supported by the NSF award CCF-2043134. Additionally, this research was supported by the Center for Nonlinear Studies and the Laboratory Directed Research and Development program of Los Alamos National Laboratory under project number 20230771DI. This material was based upon work supported by the U.S. Department of Energy, Office of Science, Office of Advanced Scientific Computing Research under Triad National Security, LLC (‘Triad’) contract grant 89233218CNA000001 [FWP: LANLE2A2].

\bibliographystyle{IEEEbib}
\bibliography{refs}

\end{document}